\documentclass[12pt]{iopart}
\usepackage{graphicx}% Include figure files
\usepackage{dcolumn}% Align table columns on decimal point
\usepackage{bm}% bold math
\usepackage{cite}

\newcommand{\beq}{\begin{equation}}
\newcommand{\eeq}{\end{equation}}
\newcommand{\bea}{\begin{eqnarray}}
\newcommand{\eea}{\end{eqnarray}}

\begin{document}

\title[]{\textsl{Ab initio} study of transport properties in defected carbon nanotubes: an O(N) approach}

\author{Blanca Biel$^1$\footnote{Present address:
CEA, LETI-MINATEC, 17 rue des Martyrs, 38054 Grenoble, Cedex 9 France}, F.J. Garc{\'\i}a-Vidal$^1$,
\'Angel Rubio$^{2,3}$ and Fernando Flores$^1$}

\address{$^1$ Departamento de F{\'\i}sica Te\'orica de la Materia Condensada,
Universidad Aut\'onoma de Madrid, E-28049 Madrid, Spain}
\address{$^2$ European Theoretical Spectroscopy Facility (ETSF), Departamento de F{\'\i}sica de Materiales , Universidad Pa{\'\i}s Vasco, Edificio Korta, Avd. Tolosa 72, 
 20018 San Sebasti\'an (Spain)}
\address{$^3$ Unidad de F{\'\i}sica de Materiales Centro Mixto CSIC-UPV/EHU and Donostia International Physics Center (DIPC), 20018 San Sebasti\'an (Spain)}
\ead{blanca.biel@cea.fr}

\begin{abstract}
A combination of \textsl{ab initio} simulations and linear-scaling Green's
functions techniques is used to analyze the transport properties
of long (up to one micron) carbon nanotubes with realistic disorder. The
energetics and the influence of single defects (mono- and di-vacancies) on the
electronic and transport properties of single-walled armchair carbon nanotubes
are analyzed as a function of the tube diameter by means of
the local orbital first-principles Fireball code. Efficient O(N)
Green's functions techniques framed within the Landauer-B\"uttiker formalism allow a statistical
study of the nanotube conductance averaged over a large sample of defected tubes
and thus extraction of the nanotubes localization length. Both the cases of zero and room temperature are addressed.
\end{abstract}
\submitto{\JPCM}
\maketitle
\normalsize
\section{Introduction}
Carbon nanotubes\cite{Iijima1991,Charlier2007} are one of the most promising materials for
future nanoelectronics due to their unique electrical transport properties
\cite{reviews}. In perfect single-walled carbon
nanotubes (SWCNTs), ballistic electron conduction has been
observed \cite{exp_ballistic}
provided inelastic processes can be neglected. In this coherent regime, however,
the presence of defects, dopants or other impurities will particularly affect
the performance of the device \cite{exp_defects},
as it will determine the transport in nanotubes
from a ballistic regime to either weak or even strong localization regimes, where the resistance will
increase exponentially with the length of the system.

To fully achieve control
on the properties of carbon nanotubes, it is thus essential to
perform accurate analyses of realistic disorder on carbon nanotubes-based devices.
Although simulation tools based on Density Functional Theory (DFT) provide great accuracy and are particularly
suitable for the study of defects in a wide range of materials,
their applicability is limited by their poor scaling, which leads to a huge computational effort when
the number of atoms in the system is larger than a few hundred. This prevents from using standard
\textsl{ab initio} techniques to analyze directly the properties of electronic devices in the mesoscopic scale. 

To overcome this limitation, we have combined the use of standard (no linear-scaling) 
first principles methods to extract the electronic properties of single defects in carbon nanotubes
with the use of O(N) Green's functions techniques to study the transport properties of long tubes
with a random distribution of those defects. This approach allows to calculate characteristic transport lengths
such as the localization length for different densities of defects, providing an interesting tool to
tune the transport properties of carbon nanotubes, for instance, by a controlled creation of the defects \cite{Gomez2005} through irradiation, and
was able to predict the existence of the Anderson localization regime in these carbon systems at room temperature
as long as the electronic phase coherence is preserved \cite{Gomez2005,Biel2005}. 
The emergence of the Anderson localization phenomenon in chemically doped carbon nanotubes has been
theoretically predicted in other works \cite{Latil2004,Avriller2006},
in which a similar approach based on a tight-binding modeling from \textsl{ab initio} calculations of single
B or N impurities was used to characterize the transition between different transport regimes. 
\section{Theoretical framework}
Our analysis of the transport properties
of the defected carbon nanotubes was separated into two steps. First, we studied the
perturbation that single, isolated defects cause --at zero temperature--
on the geometry and the
electronic properties of an otherwise ideal tube.
In a further step, a study of the transport properties of nanotubes with a more realistic
disorder has been performed, calculating the conductance along the nanotubes,
both at zero and finite temperature, for a
statistically meaningful sample of nanotubes with different concentrations of randomly distributed
defects.

Our approach combines first-principles
calculations to obtain the effective one-electron Hamiltonian
for the defected nanotubes
with non-equilibrium Green's functions techniques, which allow the
calculation of the conductance for very long --up to microns-- tubes.
\subsection{Geometry optimization}
In our calculations we have used the \textsl{ab initio}
Fireball'96 code \cite{Sankey1989,Demkov1995,Ortega1998}, that
provides fast
and accurate molecular dynamics calculations of the
geometries and electronic properties for a wide range of systems. This method
uses localized orbitals,
generated by solving the atomic problem within the
DFT-LDA and the pseudopotential approximations.
Norm-conserving
pseudopotentials are used, as well
as the Ceperly-Adler form of the exchange-correlation
potential as parametrized by Perdew and Zunger
\cite{Perdew1981}.
The forces on each atom are
calculated by means of a variation of the Hellmann–
Feynman theorem, and molecular dynamics are
used to obtain the lowest-energy atomic configuration.
A \textsl{self-consistent}
implementation
of the Harris functional allows to avoid the much time-consuming calculation
of the electronic charge density $\rho(\vec r)$,
which is substituted by the calculation of
the \textsl{orbital occupancies} $n_\mu$ of
the `fireball' orbitals $\phi_\mu$, defined by
\beq
\rho(\vec r)\,=\,\sum_\mu n_\mu|\phi_\mu(\vec r)|^2 \,\,\,\,\, .
\label{def_orb_oc}
\eeq
This approach allows us to optimize the electronic charge density $\rho$
according to the chemical environment of the atoms.
The different values of
$n_\mu$ are then obtained by imposing a self-consistent
condition on the orbital occupancies.

In the simulations with Fireball'96
an $sp^3$-basis set of orbitals was used with a cut-off
radius of 2.15 \AA.
The supercell contained 12, 10 and 6 unit cells
for the $(5,5)$, $(7,7)$ and $(10,10)$ nanotubes, respectively (the unit cell
containing 20, 28 or 40 atoms for each tube).
Four special
\textsl{k}-points uniformly distributed
along the first Brillouin zone of the nanotubes
have been included. Convergence of the calculations is achieved
for forces acting over each atom being less than 0.01 \textsl{eV/\AA}
and changes on the total energy of the system are below 10$^{-3}$ \textsl{eV};
the time step was 0.5 \textsl{fs}.
\subsection{Transport formalism}
\label{transport}
The calculation of the conductance has been performed within
the Landauer-B\"uttiker formalism, more concretely in the Keldysh
(non-equilibrium Green's functions) formalism,
which is particularly convenient when
using a localized basis set.
The differential conductance $g$ between two systems in the limit
of zero temperature and low voltages is given by \cite{Martin-Rodero1988,Mingo1996}
\beq
g(E)\,=\,\frac{4\pi e^2}{\hbar}\,Tr{[\,\hat{D}^A_{11}(E)\,\hat{T}_{12}\,
\hat{\rho}^{(0)}_{22}(E)
\,\hat{T}_{21}\,\hat{D}^R_{11}(E)\,\hat{\rho}^{(0)}_{11}(E) \,]}
\label{formula_conductancia_Bardeen}
\eeq

where \textsl{Tr[$\hat{O}$]} represents the trace of operator \textsl{$\hat{O}$};
$\hat{T}_{12}$ describes the coupling between systems 1 and 2;
$\hat{\rho}^{(0)}_{11}(E)$ and $\hat{\rho}^{(0)}_{22}(E)$
are the density of states matrices associated with the decoupled
($\hat{T}_{12}$ = 0) systems 1 and 2.
Multiple scattering effects are included by means
of the retarded and advanced denominator functions $\hat{D}$:
\bea
\hat{D}^A_{11}\,=\,[I-\,\hat{T}_{12}\,\hat{G}^{(0)A}_{22} \,\hat{T}_{21}\,\hat{G}^{(0)A}_{11}]^{-1}
\nonumber\\
\hat{D}^R_{22}\,=\,[I-\,\hat{T}_{21}\,\hat{G}^{(0)R}_{11} \,\hat{T}_{12}\,\hat{G}^{(0)R}_{22}]^{-1} \,\,\,\,\, .
\label{def_denominador}
\eea

\begin{figure*}[htbp]
\centering
\includegraphics[width=14cm,height=3cm]{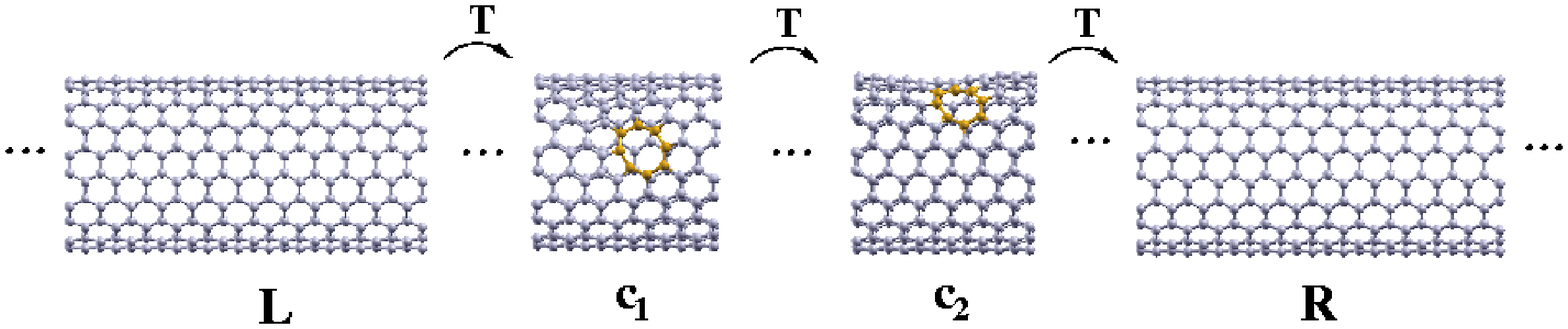}
\caption{(color online) Nanotube scheme as considered to perform
the conductance calculation:
\textsl{L} and \textsl{R} are the semi-infinite
ideal nanotubes that act as left and right electrodes;
the central region is the defected nanotube,
consisting of one or more blocks of ideal (`\dots') or defected
($c_1$, $c_2$, ...) nanotube.}
\label{nano_Dyson}
\end{figure*}

We are interested in calculating the conductance of a defected nanotube,
but we are not including in this analysis the effect of realistic contacts.
The simulation geometry will thus consist of a device region where the
nanotube containing the defects is connected to two
semi-infinite perfect tubes (\textsl{L} and \textsl{R})
that act as left and right electrodes (see Figure \ref{nano_Dyson});
the density of states of the electrodes is calculated using standard decimation techniques.
The conductance along the defected region of the nanotube is calculated by an iterative procedure,
starting on the left electrode \textsl{L}, and including at
each step a random number of ideal layers of nanotube
(shown as `\dots' in Figure \ref{nano_Dyson}) as well
as the block (supercell) of the nanotube containing the defect, and whose most stable reconstruction
has been obtained using Fireball'96
(labeled as $c_1$ and $c_2$  in the same figure).
After
the inclusion of a new defect, the conductance between the `left side' of
the system (consisting of the left lead and the
increasing
number of ideal and defected layers of nanotube) and the right electrode (which
remains invariable during the process) is calculated. In this way, we can calculate
the conductance as a function of the number of defects included
(and thus of the length of the tube) for different random configuration of defects
distributed along the nanotube with a given mean distance \textsl{d} between them.
The length \textsl{L} of the tubes is obtained from $L\,=\,N\cdot d$, where $N$ is
the number of defects in the nanotube.

The main advantage of this linear-scaling numerical procedure is the fact that the size of the matrices we need
to invert at each step does not increase with the size of the system,
but remains constant and equal to the size of the defected region of the nanotube. In this way, we can calculate 
the conductance for mesoscopic-size tubes with different concentrations of defects without a very expensive
computational effort.

The \textsl{hopping} or coupling between the layers
of nanotube is assumed to be the coupling between the bulk layers of
ideal nanotubes. For this reason, it is essential that the
number of layers included in the supercell of the \textsl{ab initio} simulations
is large enough so that the surface layers of the supercell are not being affected
by the perturbation caused by the defect
and can be regarded as
`ideal' bulk layers.

At zero temperature, the conductance
for an infinitesimal voltage is calculated just by the evaluation
of the differential conductance $g(E)$ at the Fermi level $E_f$.
However, at finite T there will be more states
accessible to the electrons and the calculation of the
conductance will be performed by means of the following
expression \cite{Buttiker1985}:
\beq
G\,=\,\int_{-\infty}^{+\infty} (-\frac{d\,f_T(E)}{dE})\, g(E)\, dE  \,\,\,\,\, ,
\label{G_temp_finita}
\eeq
where
\beq
f_T(E)\,=\,\frac{1}{e^{\frac{E-E_F}{k_BT}}+1}
\eeq
is the Fermi distribution function.
In this way, the total conductance at a given temperature is calculated as the total contribution
from all the electronic states included in the thermal energy window opened by the finite temperature.
\section{Analysis of a single defect}
Theoretical simulations \cite{Krash2001,Krash2002}
suggest that Ar$^+$ irradiation on carbon nanotubes will mainly create monovacancies and
divacancies along the nanotube. We have therefore studied the monovacancy and the two possible
orientations of the divacancy, which we will refer to as vertical (if the orientation of the chain 
of missing atoms is perpendicular to the nanotube direction)
and lateral (if the orientation is partially parallel to the tube axis).
\subsection{Monovacancies}
The structural and electronic properties of vacancies-related defects in CNTs and other graphene-based materials
have been extensively analyzed in the literature (see, for instance, \cite{Ajayan1998,Ewels2002,Lu2004,Krash2006,Amorim2007,Amara2007}).
The extraction of one single atom in the nanotube leads to a metastable state,
that propitiates recombination of nearest neighbors to fill the empty space left by the vacancy.
\subsubsection{Effect on the geometry}

\begin{figure}[htbp]
\centering
\includegraphics[width=3.8cm,height=2.3cm]{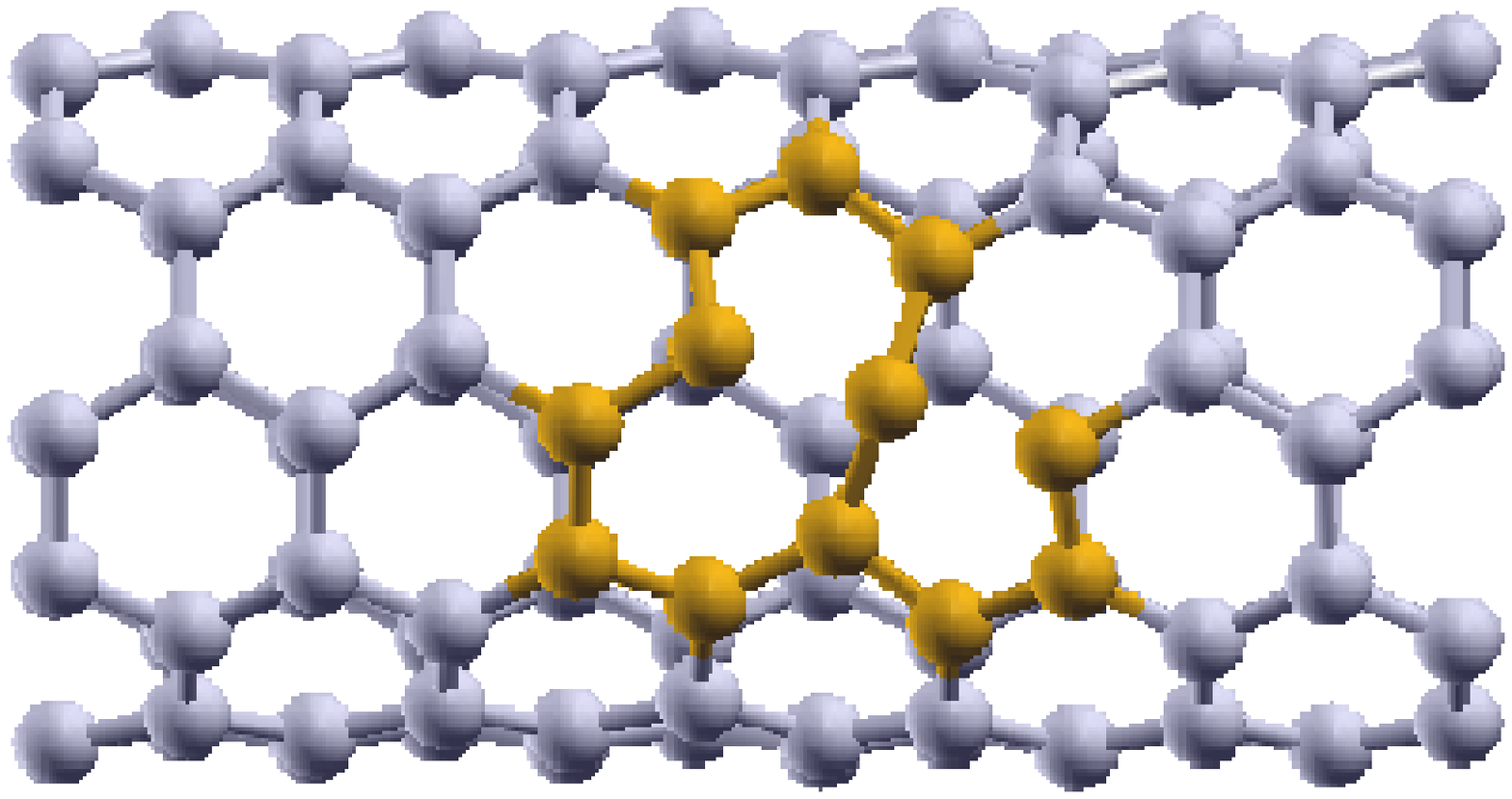}
\hskip0.8cm
\includegraphics[width=4.0cm,height=2.6cm]{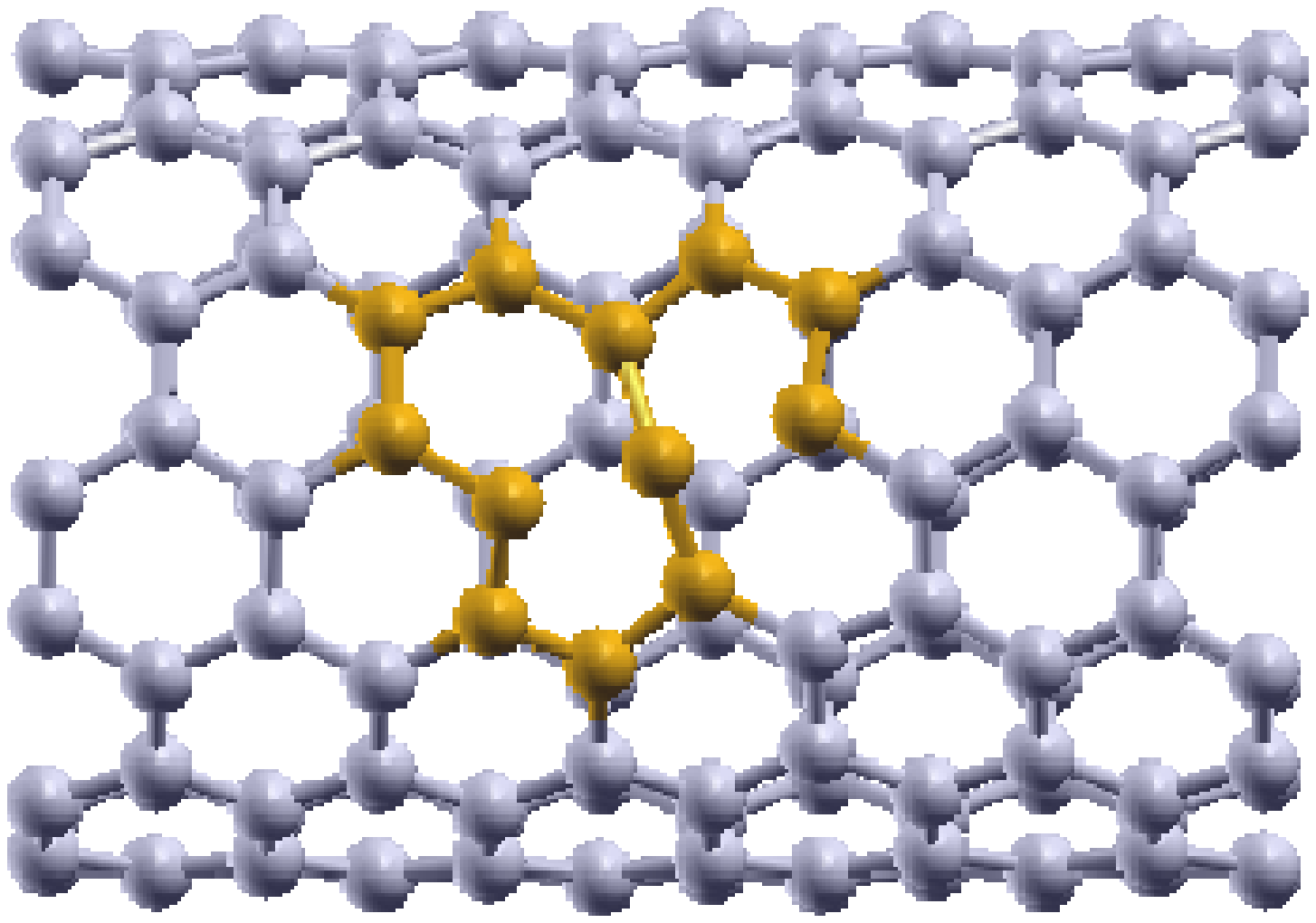}

\vskip1.5cm
\includegraphics[width=4.2cm,height=4.2cm]{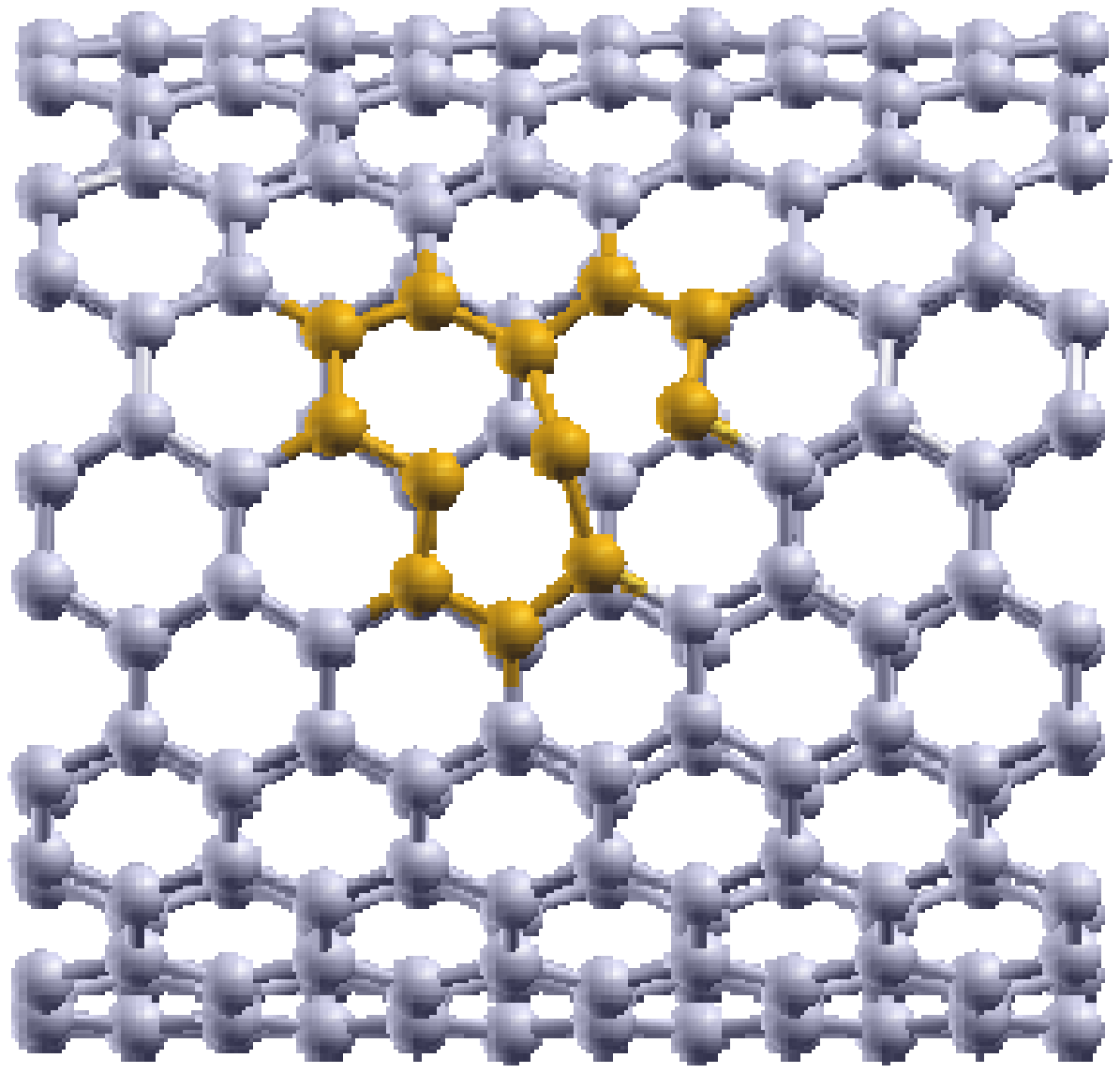}
\vskip0.5cm
\caption{(color online) Relaxed geometries of the (5,5) (top left),
(7,7) (top right) and (10,10) (bottom) SWCNTs with a single
monovacancy. Only the area around the defect (6 layers) is shown.}
\label{vac_geom}
\end{figure}

After the relaxation, we found that new bonds are formed between
the neighbors of the extracted atom, favoring three-atoms against two-atoms coordination.
This well-known recombination of atoms is found in the three types of armchair CNTs analyzed in this work.

The tube diameter is also affected by the extraction of the single atom.
It decreases in the zone of the defect and reaches its maximum value at a distance
of $\approx$ 5 {\AA} from the defect, recovering the diameter of the ideal tube
at the surface layer of the supercells used in our simulations.
Table \ref{table_vac} shows diameter ($d$) for ideal tubes
as well as minimum and maximum diameter for defected ones.
\subsubsection{Energetics}

The analysis of the energetics of the monovacancy has been performed for each CNT
by means of the following difference of total energies, that allow us to compare
total energies for systems containing a different number of atoms:
\beq
{\Delta E}_{monovac}\,\equiv|\,E_{ideal}\,-\,(E_{monovac}\,+\,E_{atom})\,| \,\,\,\,\, ,
\eeq
where $E_{ideal}$ is the total energy for an ideal tube calculated for a supercell containing
the same number of layers than the supercell for the defected tube,
$E_{monovac}$ is the total energy of the relaxed nanotube containing the monovacancy,
and $E_{{a}tom}$ is the energy for the free atom. ${\Delta E}_{monovac}$ for each tube is also shown in
Table \ref{table_vac}.
\begin{table*}
\center
\caption{\label{table_vac} Maximum and minimum diameters for $(5,5)$, $(7,7)$ and $(10,10)$
nanotubes with a single monovacancy and comparison of ${\Delta E}_{monovac}$
for the relaxed geometries.}
\vskip0.5cm
\begin{tabular}{cccccc}
\hline
CNT & $d_{ideal}$ (\AA) & $d_{min}$ (\AA) & $d_{max}$ (\AA) & ${\Delta E}_{monovac}$ (eV)\\
\hline
$(5,5)$&6.97&6.56&7.34&16.62\\
$(7,7)$&9.76&9.36&10.20&17.41\\
$(10,10)$&13.89&13.86&14.76&17.42\\
\hline
\end{tabular}
\end{table*}
\subsection{Divacancies}
\begin{figure}[htbp]
\centering
\includegraphics[width=3.8cm,height=2.3cm]{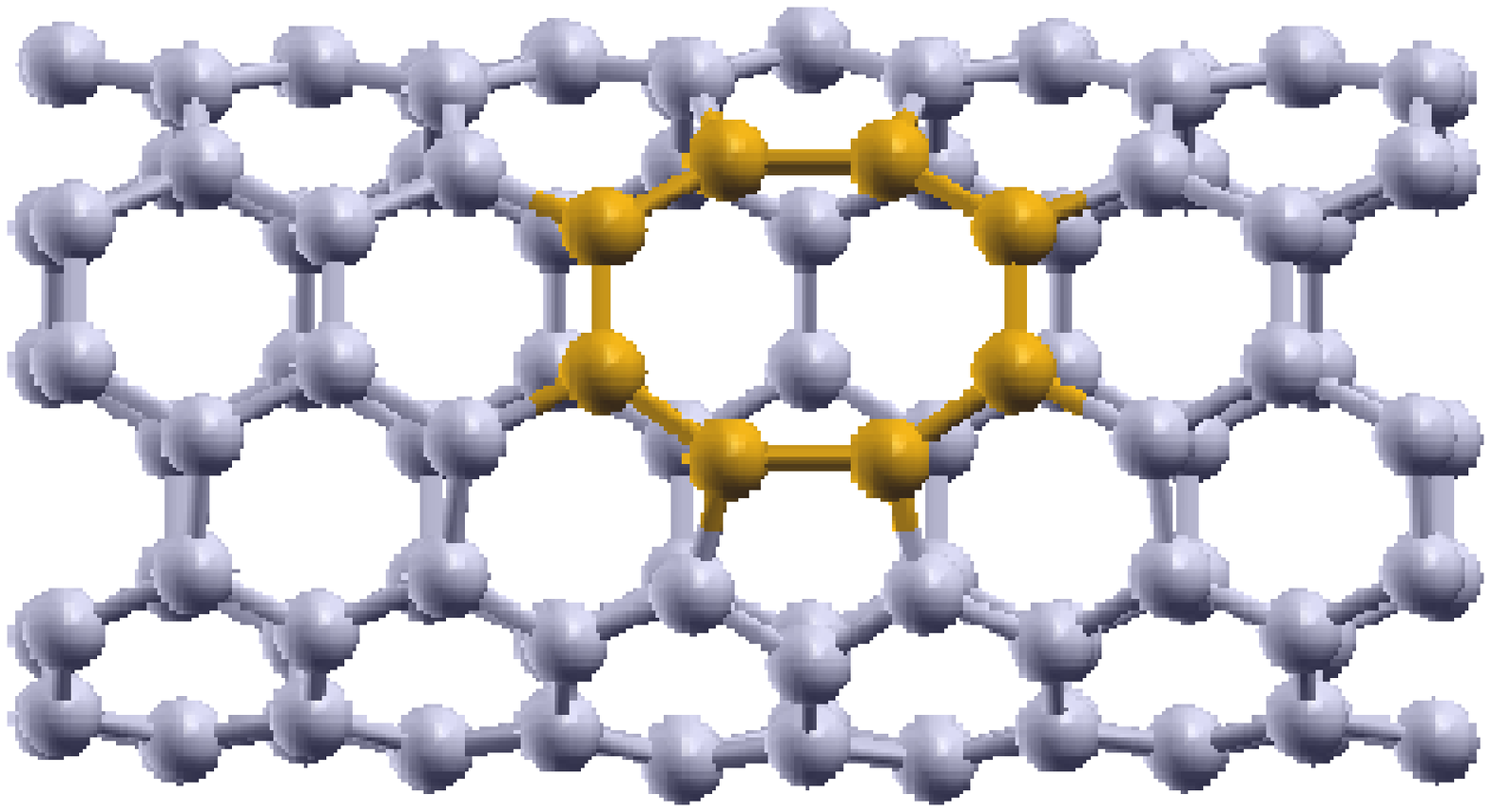}
\hskip0.5cm
\includegraphics[width=4cm,height=2.6cm]{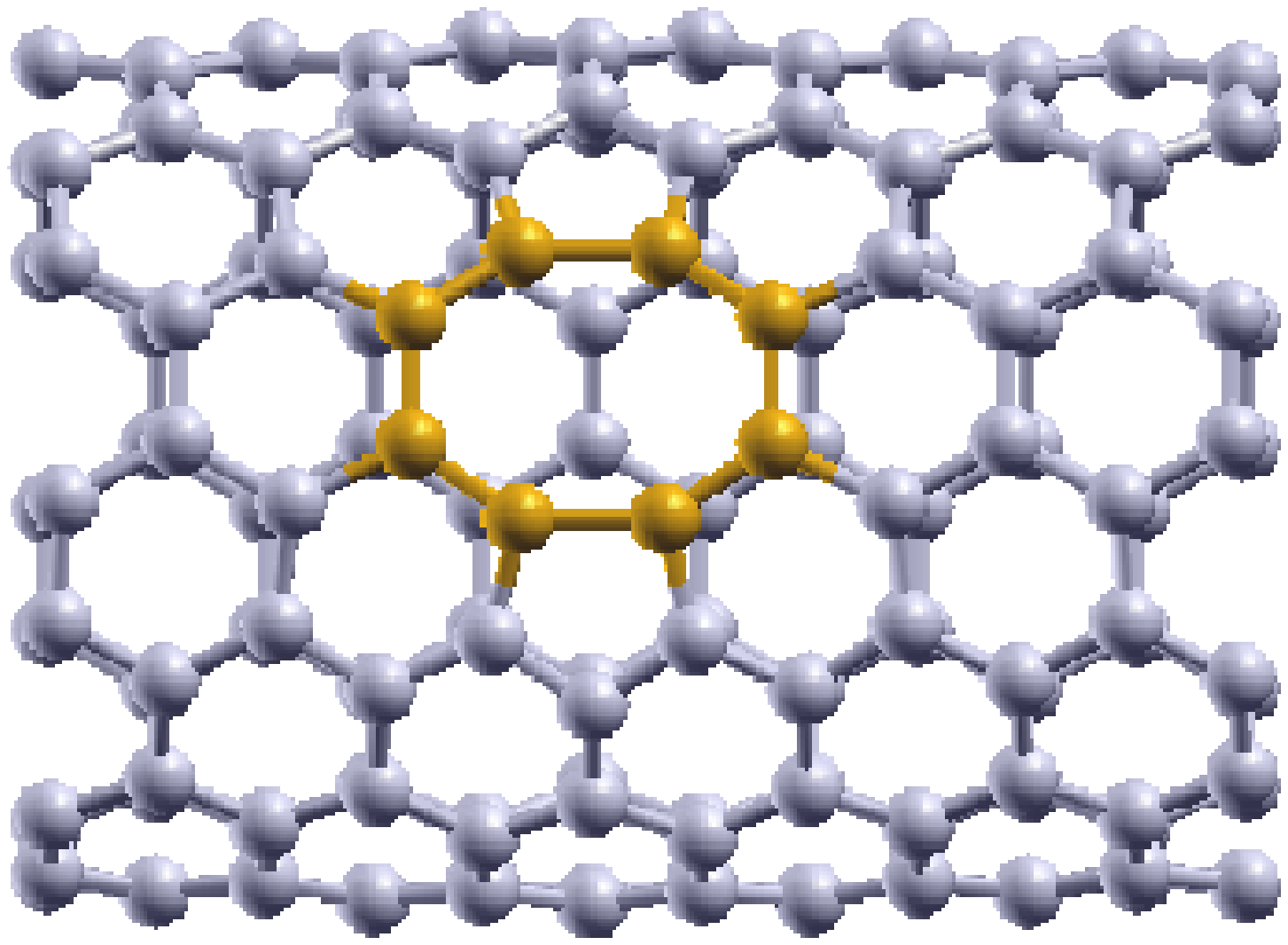}

\vskip1.5cm
\includegraphics[width=4.2cm,height=4.2cm]{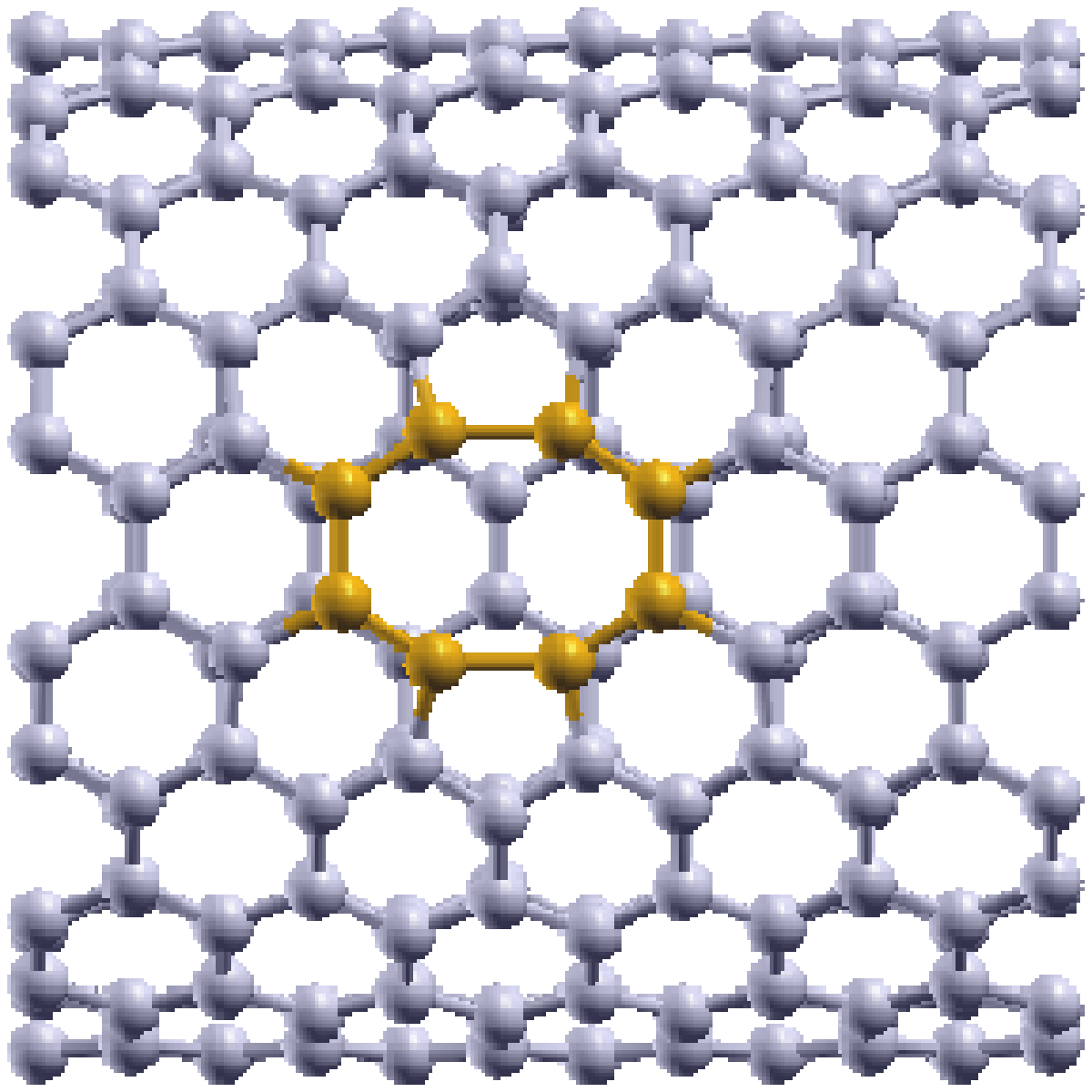}
\vskip0.5cm
\caption{(color online) Relaxed geometries of the (5,5) (top left),
(7,7) (top right) and (10,10) (bottom) SWCNTs with a vertical
divacancy. Only the area around the defect (6 layers) is shown.}
\label{vert_geom}
\end{figure}
\begin{figure}[htbp]
\centering
\includegraphics[width=3.8cm,height=2.3cm]{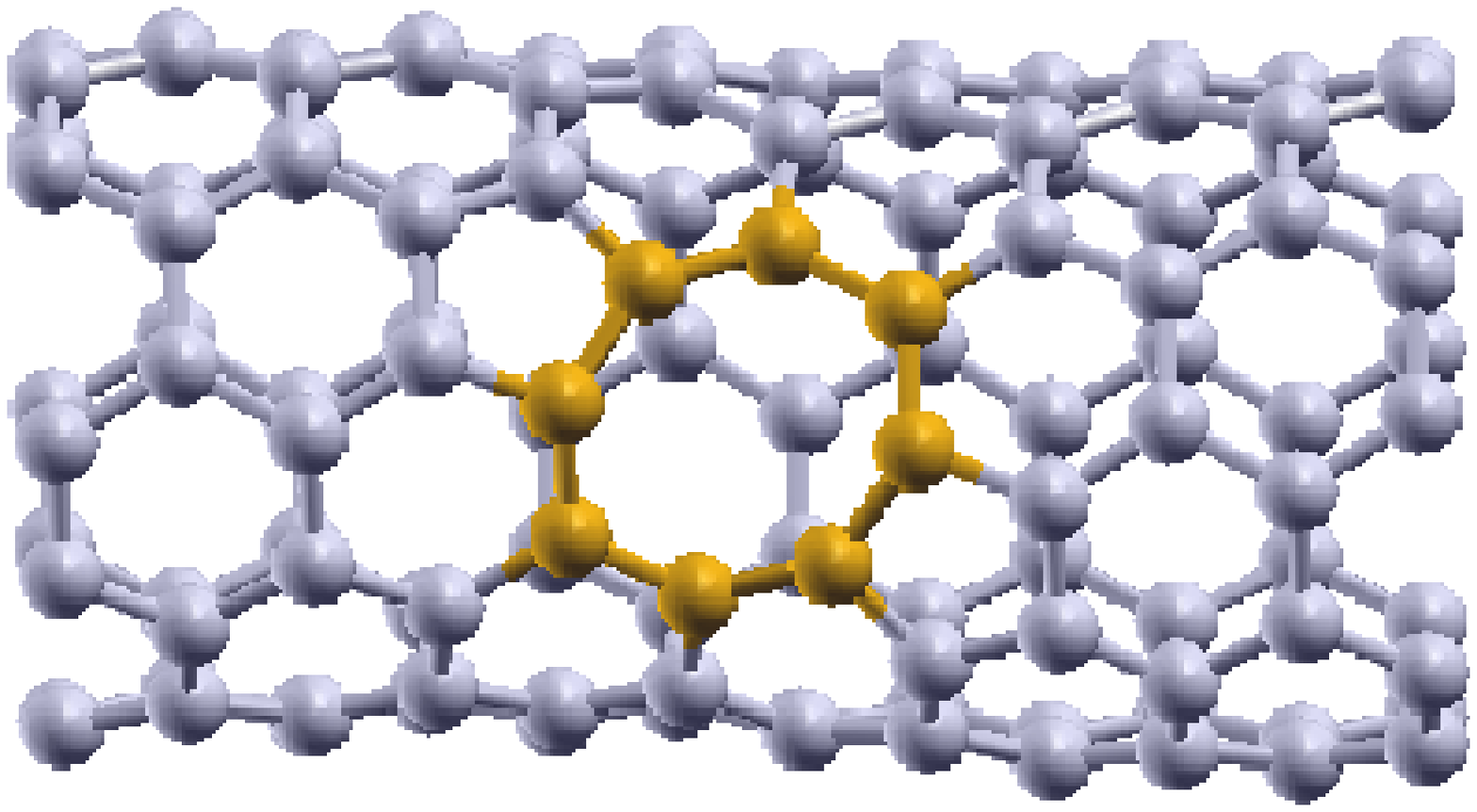}
\hskip0.5cm
\includegraphics[width=4cm,height=2.6cm]{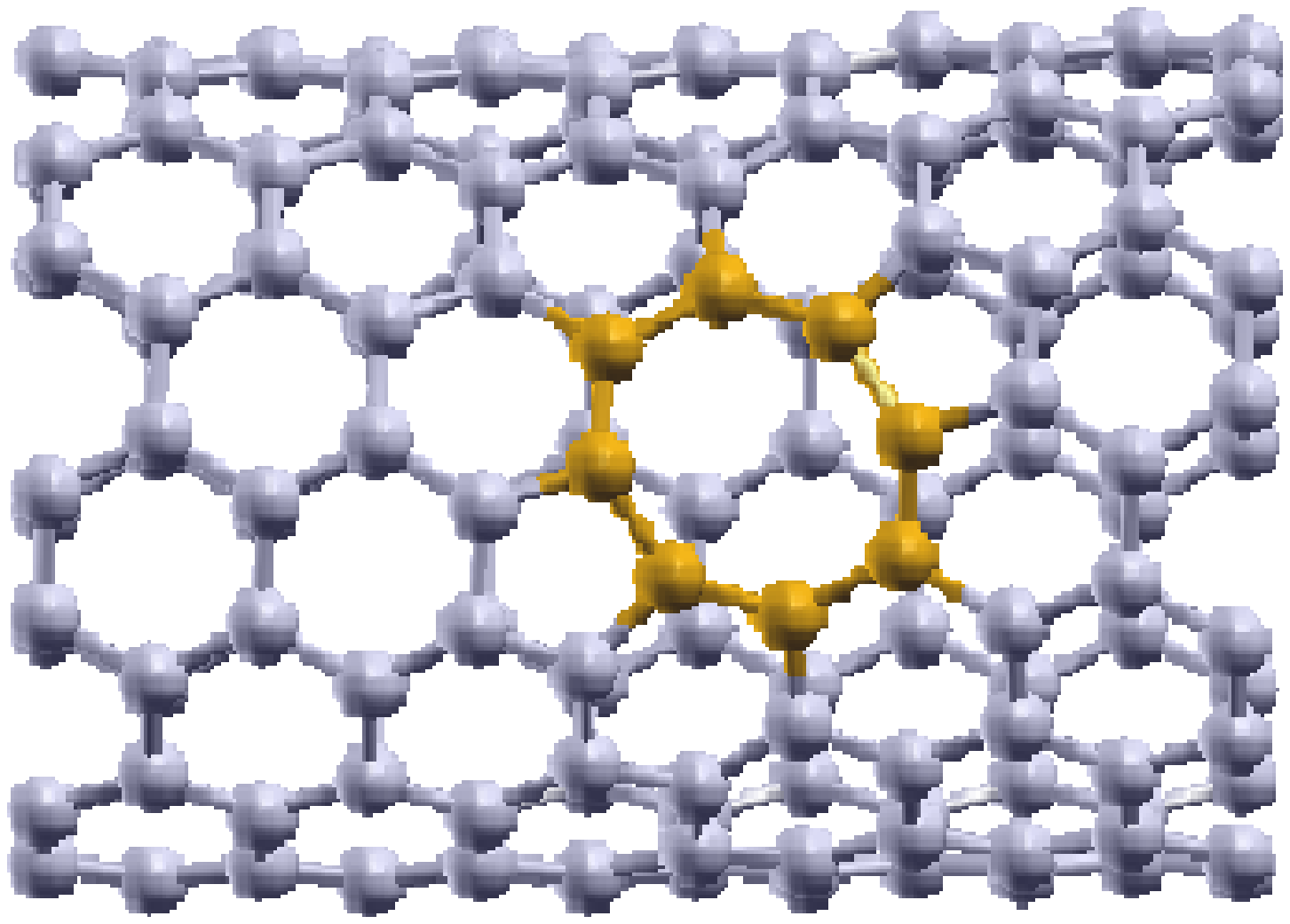}

\vskip1.5cm
\includegraphics[width=4.2cm,height=4.2cm]{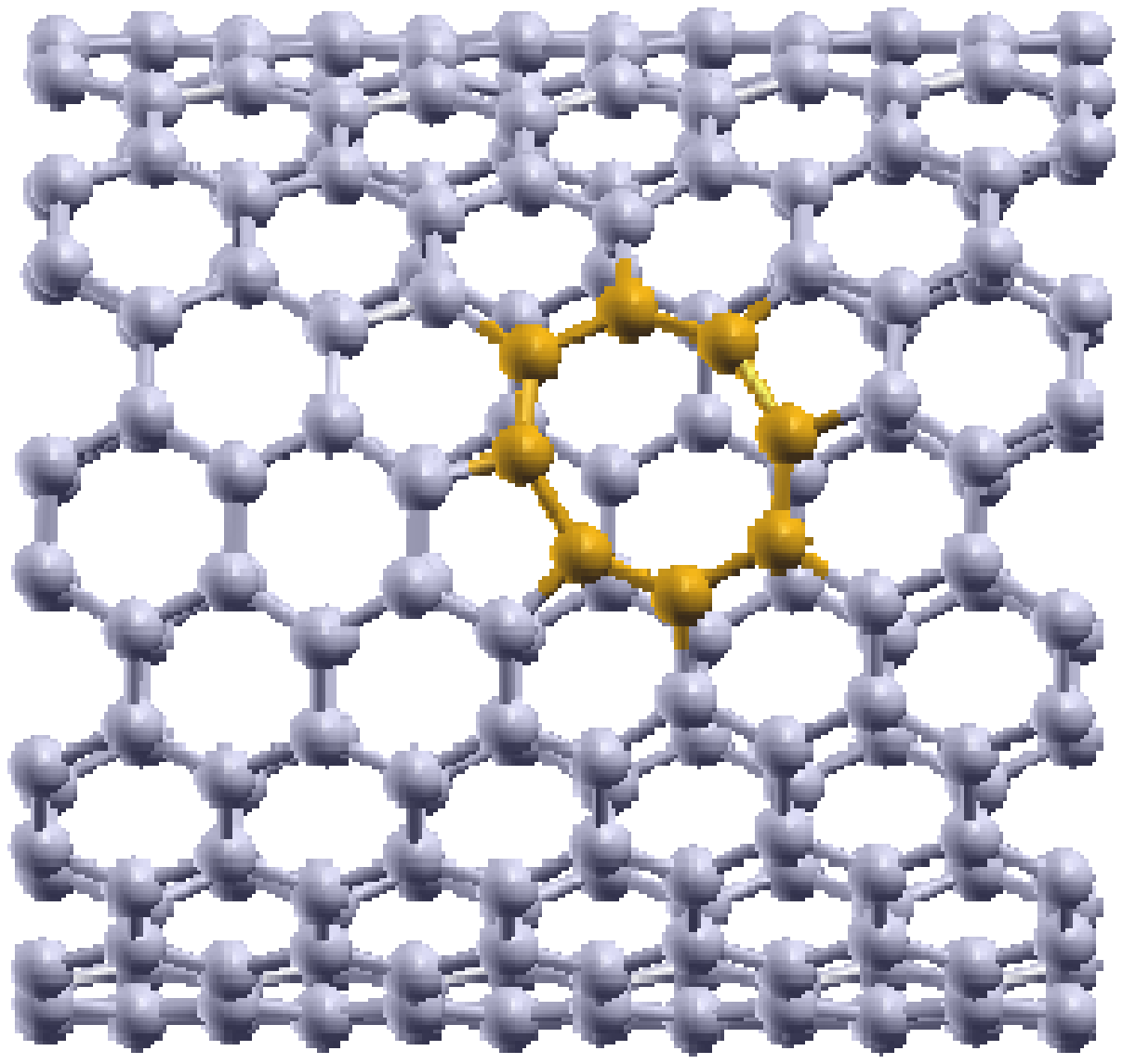}
\caption{(color online) Relaxed geometries of the (5,5) (top left),
(7,7) (top right) and (10,10) (bottom) SWCNTs with a lateral
divacancy. Only the area around the defect (6 layers) is shown.}
\label{lat_geom}
\end{figure}

Divacancies are created extracting a second atom form the nanotube, leading to two possible
orientations of the divacancy depending on which atom is extracted:
removing the atom placed
just above (or below) the first atom extracted will create a vertical divacancy (Figure \ref{vert_geom}), whereas
the lateral divacancy is formed by the extraction of the atom located to one side of the monovacancy (Figure \ref{lat_geom}).

We have found a similar reconstructed geometry for the three CNTs analyzed: atomic relaxation
leads to a new geometry formed by two pentagons and an octagon.
As for the monovacancy case, the diameter decreases locally at the zone containing the defect, but the effect
is much more pronounced for the case of the lateral divacancy (see Table \ref{table_div}).
\subsubsection{Effect on the geometry}

\vskip0.5cm
\subsubsection{Energetics}

The energy formation of the divacancies has been calculated in a similar way to the
one performed in the case of the monovacancy. Now,
\beq
{\Delta E}_{divac}\,\equiv|\,E_{ideal}\,-\,(E_{divac}\,+\,2\times\,E_{atom})\,| \,\,\,\,\, .
\eeq
According to the calculated energy formation, lateral divacancies are about 2 eV more stable
than vertical ones for all the analyzed nanotubes. 

We can also estimate the energy for an ideal system with two monovacancies infinitely
separated using
\beq
E_{2\,vac}\,=\,E_{ideal}\,-\,2(E_{ideal} - E_{vac}) \,\,\,\,\, .
\eeq
Comparing this energy difference
${\Delta E}_{divac}$ with the previously calculated ${\Delta E}_{monovac}$
\beq
\Delta \,\equiv|\,{\Delta E}_{divac}\,-\,2\times {\Delta E}_{monovac}| \,\,\,\,\, ,
\eeq
we have found that formation of a lateral divacancy is
between 6 and 7 eV more stable, for all CNTs studied, than formation
of two monovacancies. Energy formation of vertical divacancies is for all tubes
between 4 and 5 eV smaller than that of the lateral one when compared to the case of two isolated monovacancies.
Thus, lateral divacancies formation on CNTs is preferred over that of vertical ones,
and especially than formation of isolated monovacancies.
Results are summarized in Table \ref{table_div}.
\begin{table*}
\center
\caption{\label{table_div} Maximum and minimum diameters for $(5,5)$, $(7,7)$ and $(10,10)$ nanotubes with a vertical and
lateral divacancy and comparison of ${\Delta}$
for the relaxed geometries. }
\vskip0.5cm
\begin{tabular}{lccccc}
\hline
CNT & $d_{ideal}$ (\AA) & $d_{min}$ (\AA) & $d_{max}$ (\AA) & ${\Delta}$ (eV)\\
\hline
$(5,5)$ vertical divacancy &6.97&6.99&7.19&4.3\\
$(5,5)$ lateral divacancy &6.97&6.64&7.28&6.5\\
\hline
$(7,7)$ vertical divacancy &9.76&9.72&9.99&4.8\\
$(7,7)$ lateral divacancy &9.76&9.36&10.02&6.7\\
\hline
$(10,10)$ vertical divacancy&13.89&13.88&14.07&4.2\\
$(10,10)$ lateral divacancy&13.89&13.72&14.65&6.7\\
\hline
\end{tabular}
\end{table*}
\subsection{Scattering of a single defect at zero temperature}
\normalsize Although a large number of theoretical works have been performed to analyze
the scattering produced by single defects (either topological
or substitutional impurities),
it is worth mentioning that a full relaxation of the atomic region
around the defect is particularly important in the case of vacancies, where
the energetically more stable reconstructed vacancy will present a
different electronic structure than the ideal (non-reconstructed) one, leading to
a very different behavior of the conductance and other transport properties,
especially for energies close to the Fermi level.

\begin{figure}[htbp]
\centering
\includegraphics[width=6cm,height=3.2cm]{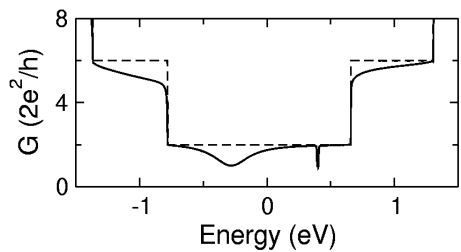}
\caption{Conductance as a function of energy for the (10,10) SWCNT with a single monovacancy, before relaxation.}
\label{vac_unrelaxed_G}
\end{figure}

\begin{figure}[htbp]
\vskip0.5cm
\centering
\includegraphics[width=6cm,height=3.2cm]{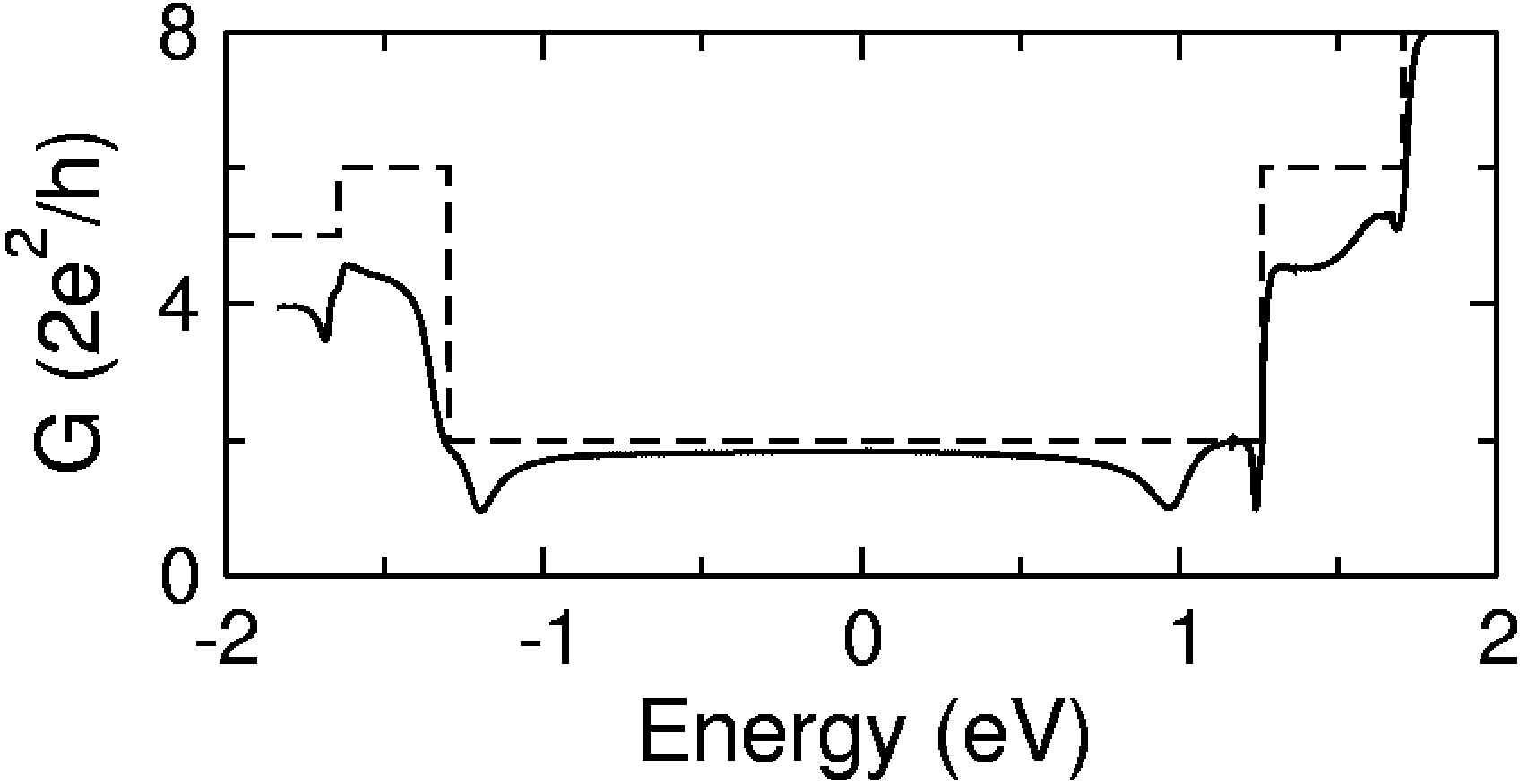}
\hskip0.5cm
\includegraphics[width=6cm,height=3.2cm]{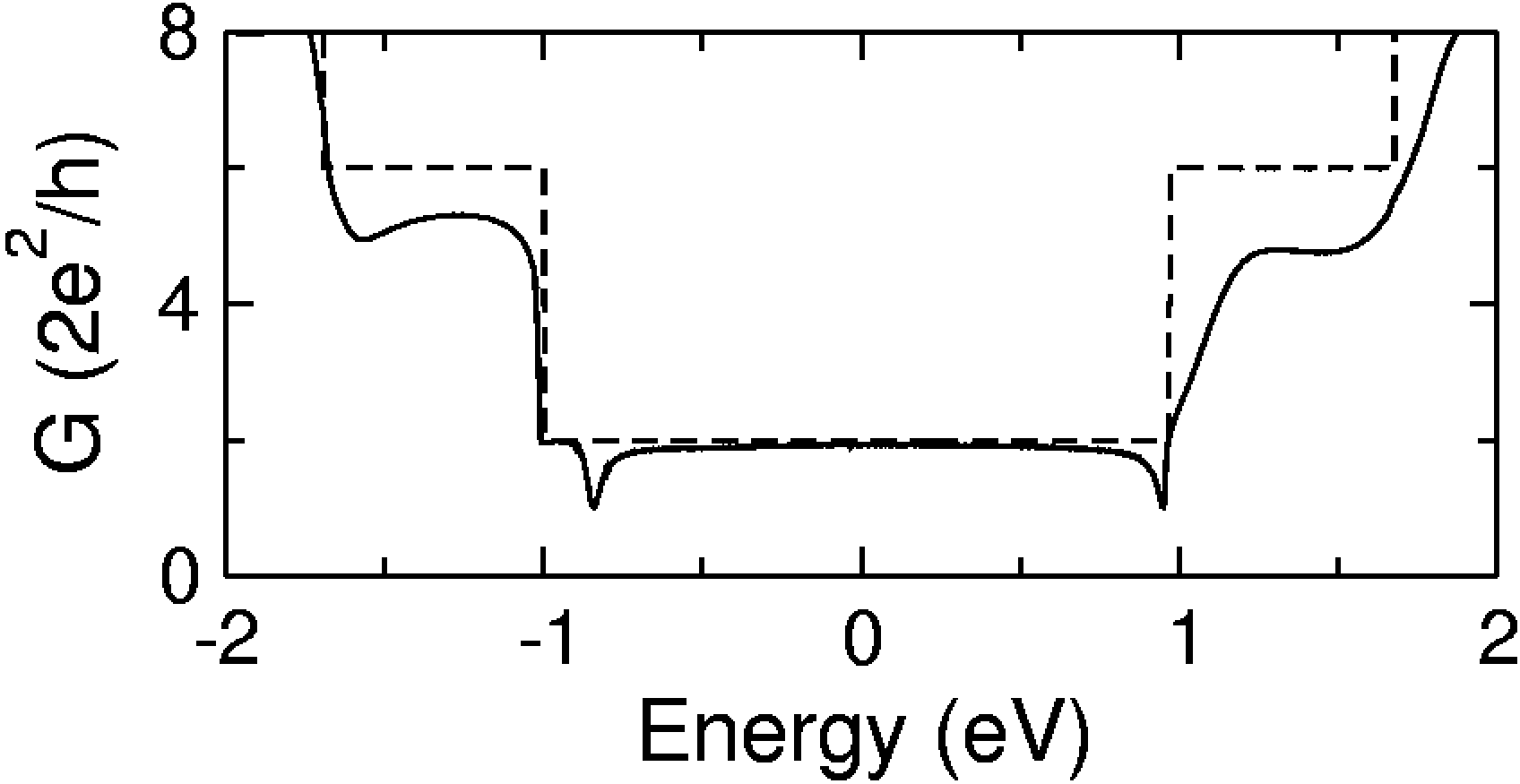}
\vskip0.8cm
\includegraphics[width=6cm,height=3.2cm]{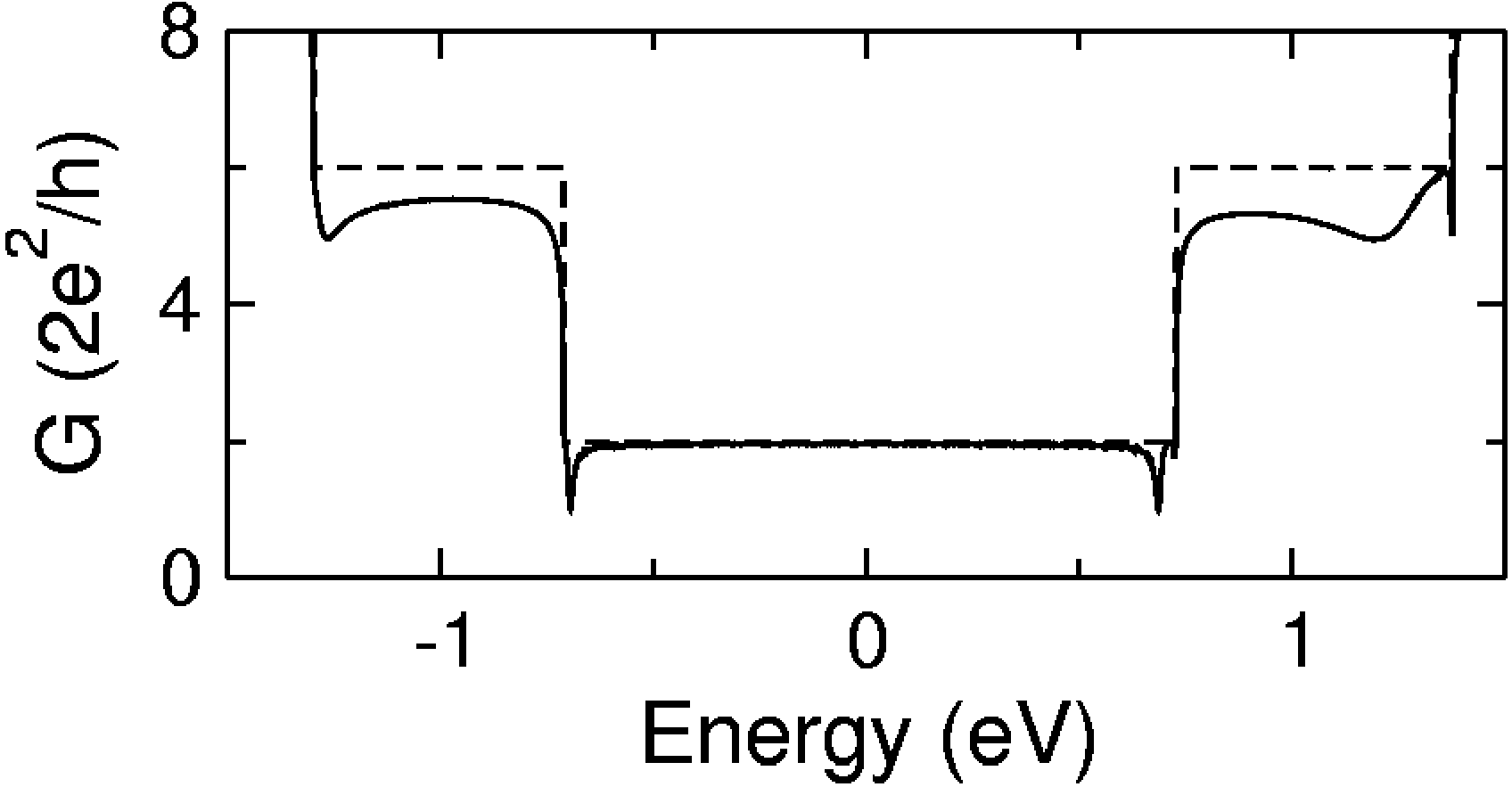}
\caption{Conductance as a function of energy for the (5,5) (top left),
(7,7) (top right) and (10,10) (bottom) SWCNTs with a single monovacancy after relaxation.}
\label{vac_G}
\end{figure}

In the case of a $\pi-\pi$ tight-binding model \cite{Chico1996}, a monovacancy in an armchair CNT
will produce a drop of almost one conductance quantum, G$_0=2e^2/h$, right at the Fermi level,
with the electron-hole symmetry fully preserved. The conductance of a single ideal monovacancy,
that is, in the metastable non-reconstructed state, will as well lead to a drop of almost $2e^2/h$, not at the Fermi level but
somewhat below at the valence band, as shown in Figure \ref{vac_unrelaxed_G} (where the Fermi energy has been set to zero).
This result agrees with previous \textsl{ab initio} calculations \cite{Choi2000,Son2007}.
\vskip0.5cm
However, the situation is very different in the case of the most stable reconstruction. Figure
\ref{vac_G} shows the conductance as a function of the energy for
the $(5,5)$, $(7,7)$ and $(10,10)$ nanotubes with a single monovacancy after relaxation.
Rearrangement of atoms has propitiated the saturation of dangling bonds, and the reconstructed nanotube
presents a conductance much more similar to the ideal case for low energy than that of the metastable monovacancy.
Particularly, there is a very small drop in the conductance for energies in the vicinity of the
Fermi level, and the narrow peaks in the conduction band due to the unsaturated $\sigma$ bonds
that are present for the non-reconstructed geometry
have disappeared.
As expected, the drop in the conductance decreases with increasing diameter of the nanotubes,
being practically negligible for the $(10,10)$ tube. This is in good agreement with previous calculations \cite{Brandbyge2002,Son2007}

The drop of the conductance is much larger for both orientations of the divacancy. In all cases a maximum drop
of about one conductance quantum is found for the two orientations of the divacancy.
This is also in agreement with previous works \cite{Amorim2007}.

\begin{figure}[htbp]
\vskip0.5cm
\centering
\includegraphics[width=6cm,height=3.2cm]{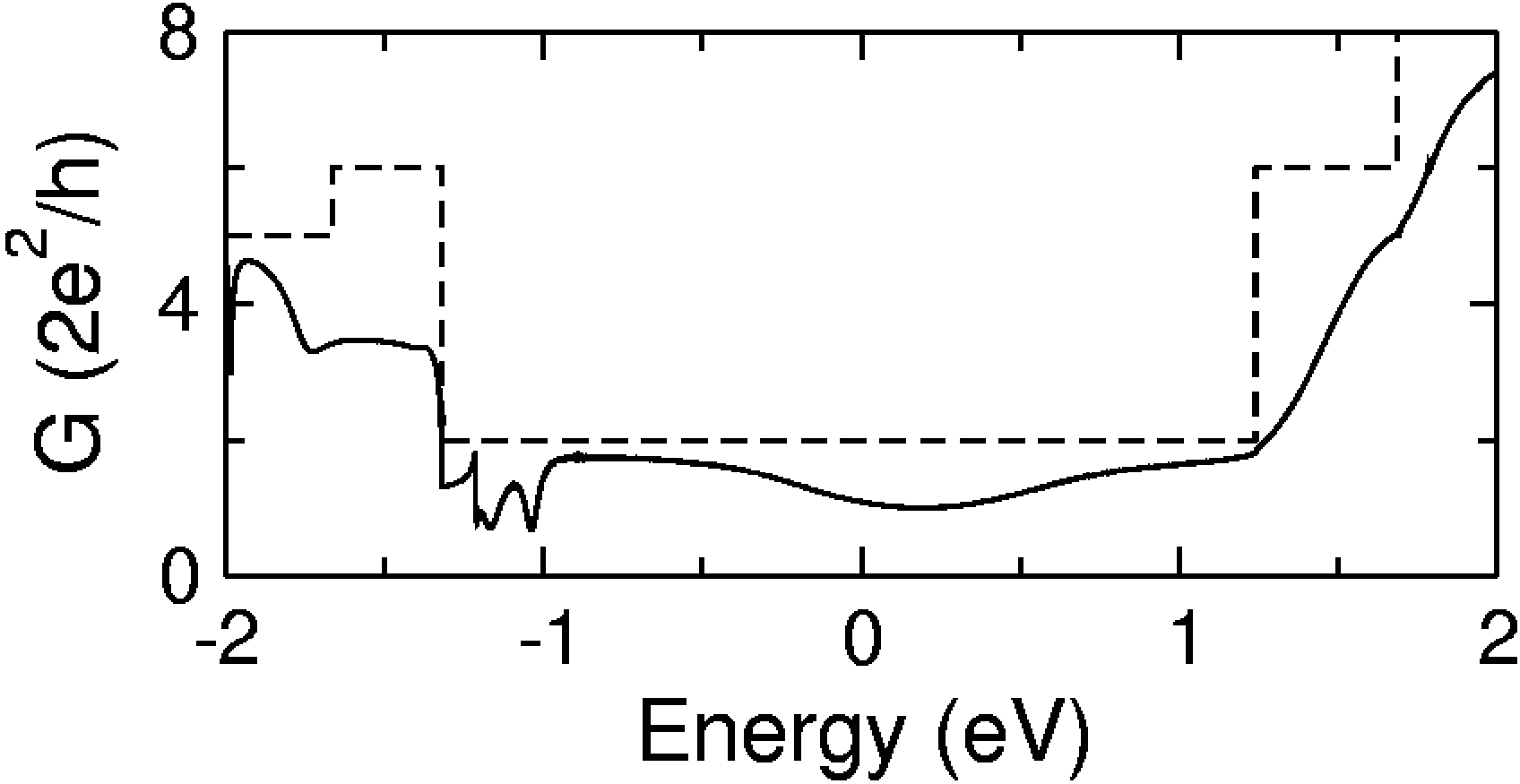}
\hskip0.5cm
\includegraphics[width=6cm,height=3.2cm]{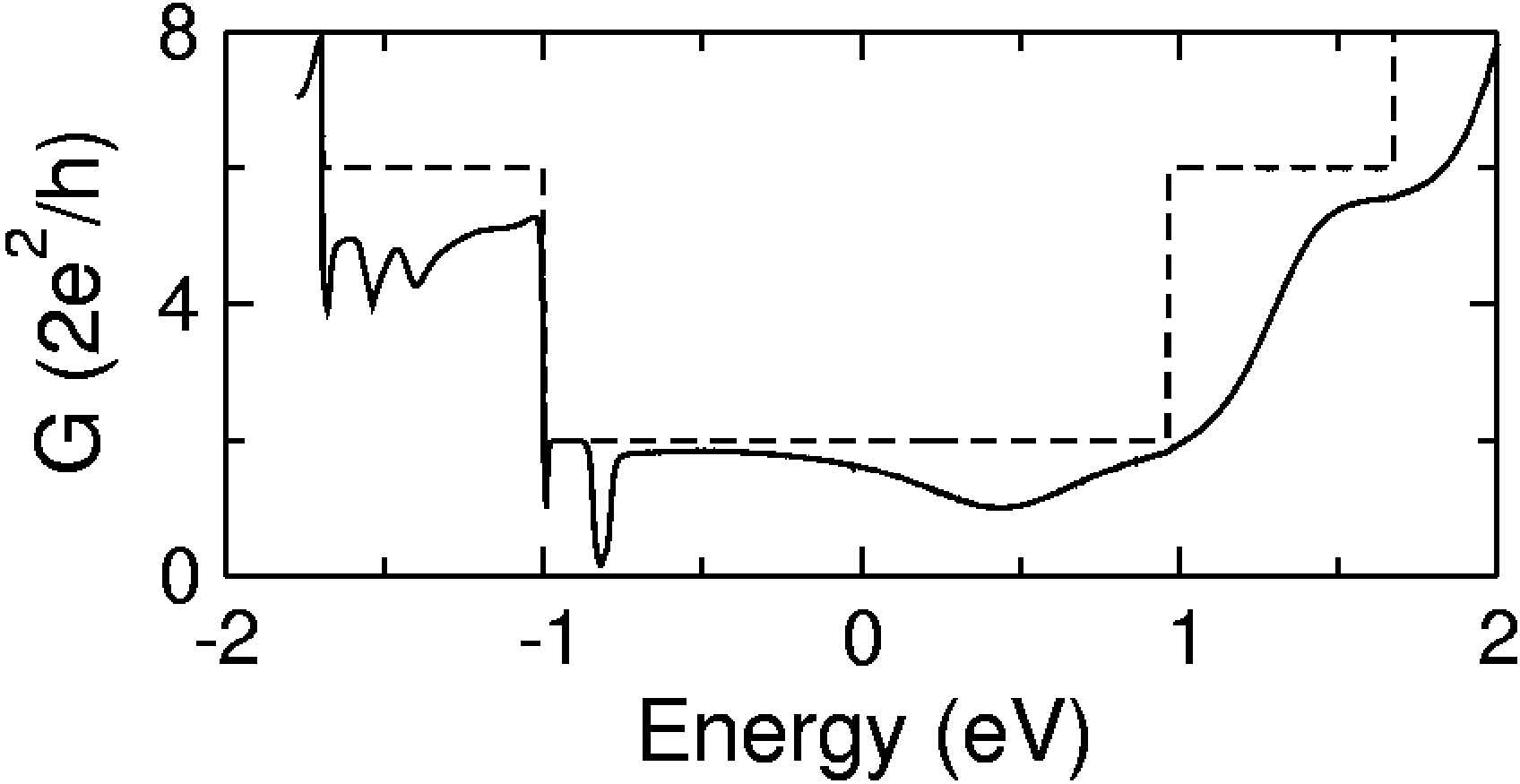}
\vskip0.8cm
\includegraphics[width=6cm,height=3.2cm]{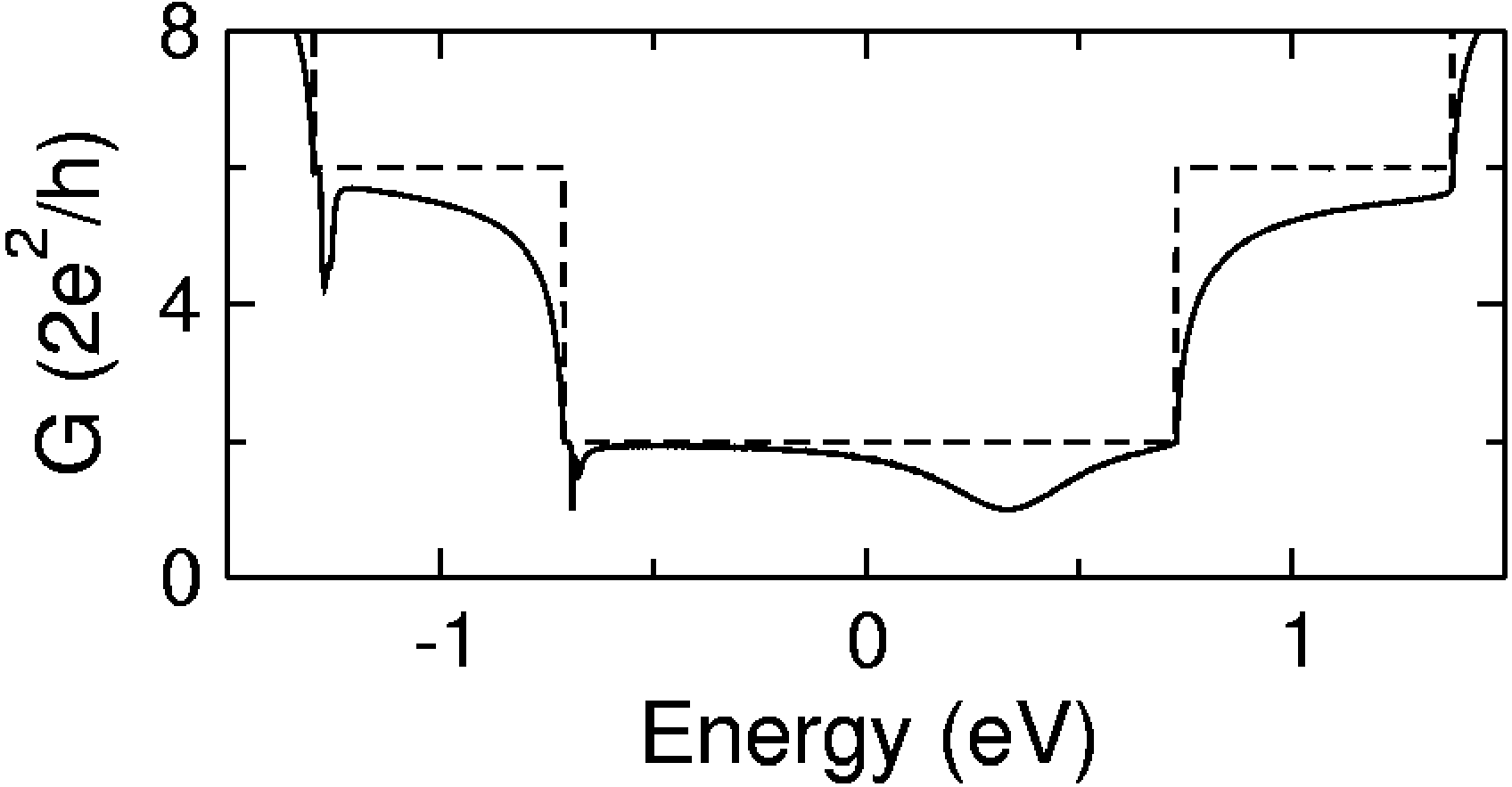}
\caption{Conductance as a function of energy for the (5,5) (top left),
(7,7) (top right) and (10,10) (bottom) SWCNTs with a single vertical
divacancy after relaxation.}
\label{vert_G}
\end{figure}
\vskip0.8cm
\begin{figure}[htbp]
\centering
\includegraphics[width=6cm,height=3.2cm]{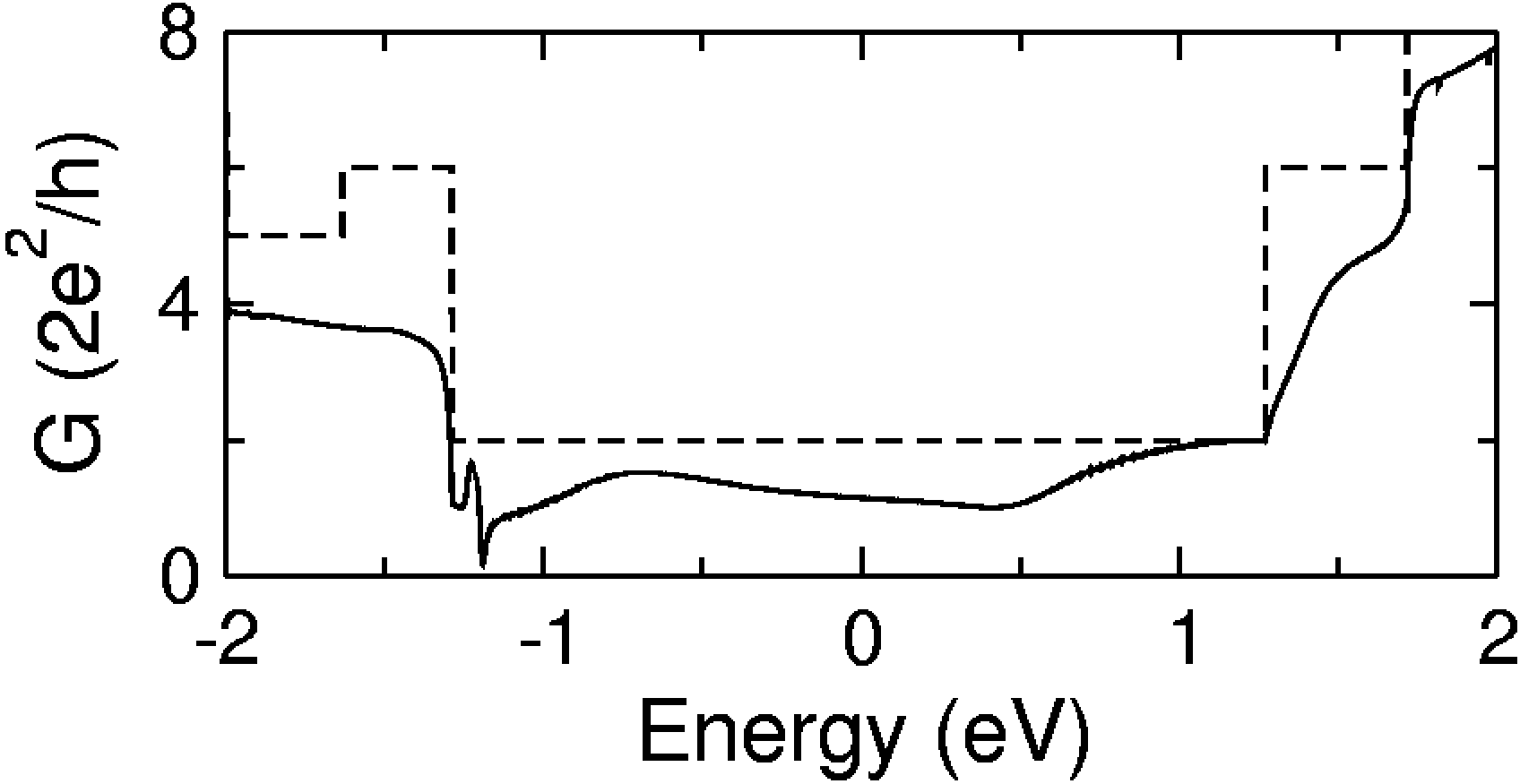}
\hskip0.5cm
\includegraphics[width=6cm,height=3.2cm]{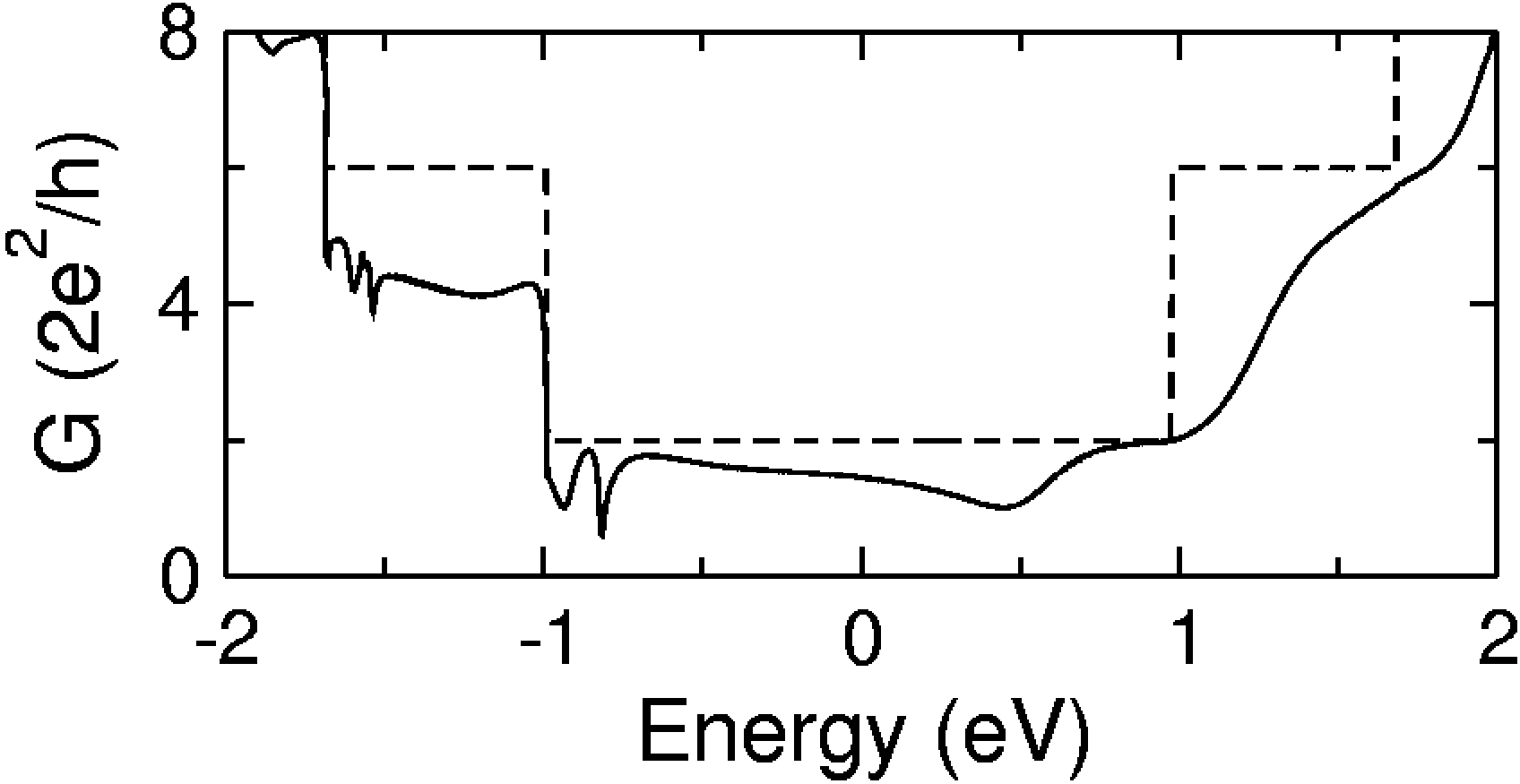}
\vskip0.8cm
\includegraphics[width=6cm,height=3.2cm]{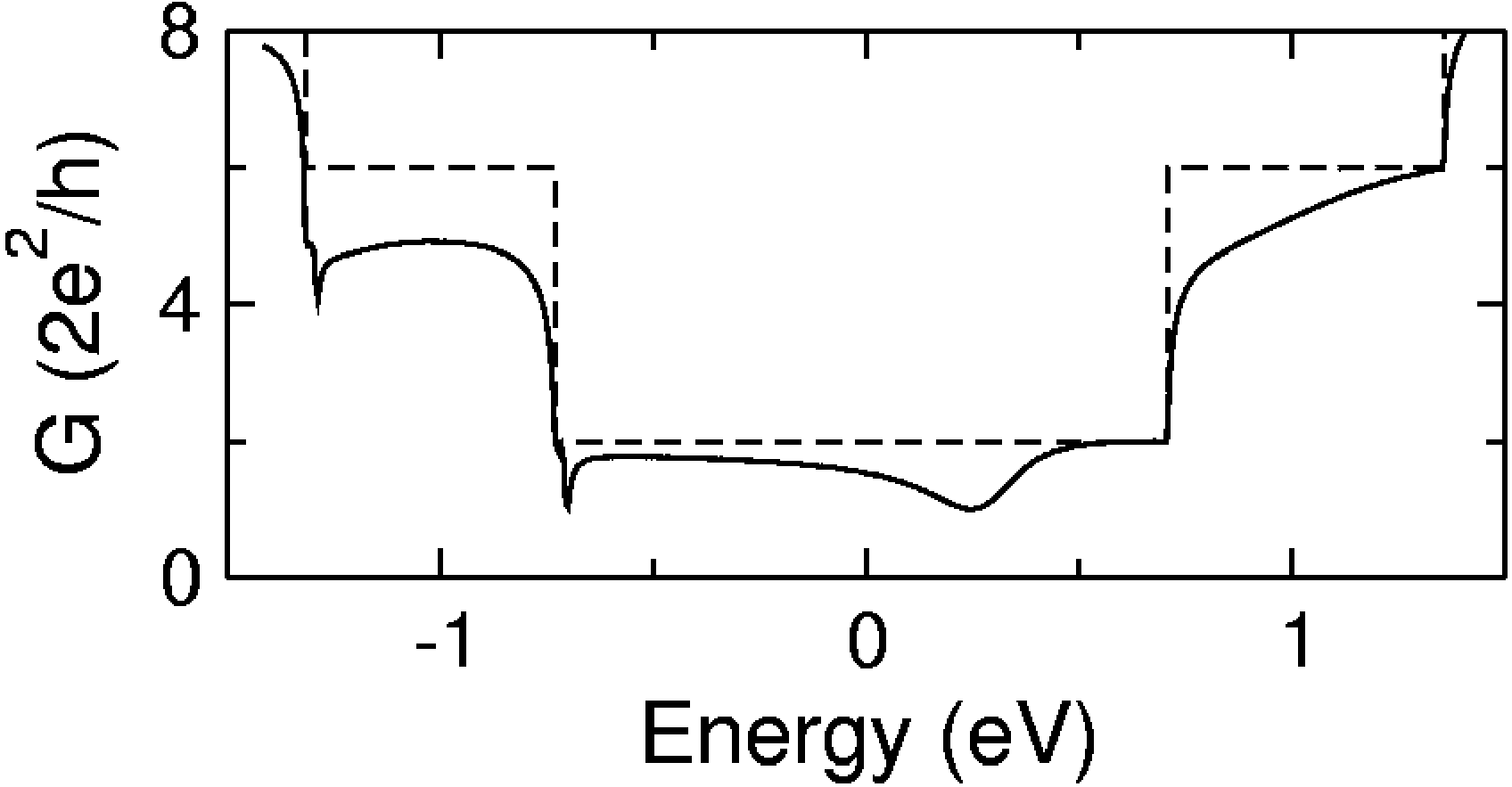}
\caption{Conductance as a function of energy for the (5,5) (top left),
(7,7) (top right) and (10,10) (bottom) SWCNTs with a single
lateral divacancy after relaxation.}
\label{lat_G}
\end{figure}

As lateral divacancies are much more stable than the vertical ones, and due to small drop
in the conductance caused by monovacancies,
the results we present
include only lateral divacancies.
\section{Statistical analysis}
Once the transport properties of the isolated defects in the infinite nanotube have been characterized,
we are interested in calculating the conductance along a more realistic nanotube in which defects can
be randomly distributed along and around the axis of the tube.
Lateral divacancies are distributed along the tube with the distance between consecutive defects
presenting a uniform random distribution between 0 and
2\textsl{d} --\textsl{d} thus being the mean distance between defects for a particular defect density.
The calculation of the resistance (as the inverse of the differential conductance)
for each particular random distribution of defects has been performed
as explained in Section \ref{transport}.

\begin{figure}[htbp]
\centering
\includegraphics[width=7cm,height=6.5cm]{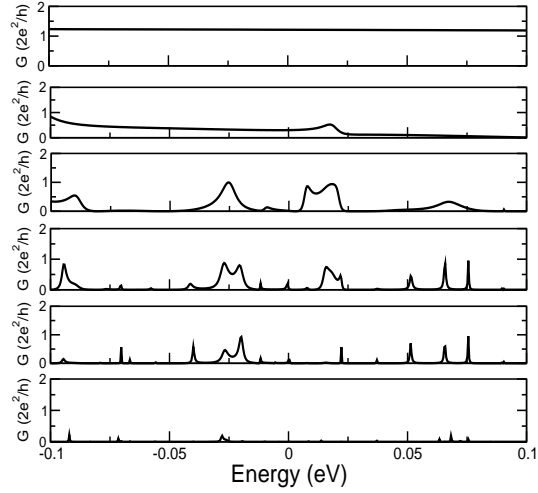}
\caption{Conductance as a function of energy for the $(5,5)$ nanotube
and a mean distance between defects of \textsl{d} = 10.1 nm for a different number, N,
of defects: N= 3, 7, 10, 15 and 20 (from top to bottom). }
\label{GvsE}
\end{figure}
\begin{figure}[htbp]
\vskip0.5cm
\centering
\includegraphics[width=8cm,height=5cm]{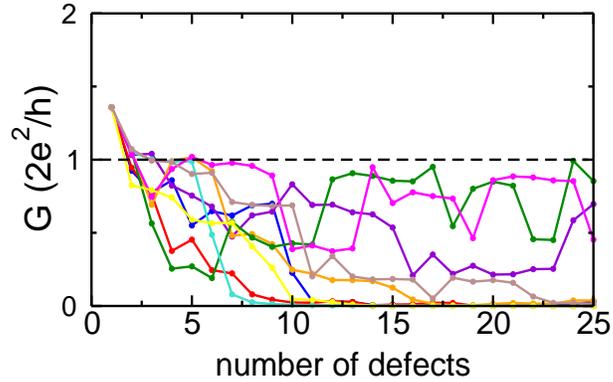}
\caption{(color online) Differential conductance at the Fermi level as a function of the number of defects,
N, for \textsl{d} = 75.5 nm for the $(10,10)$ nanotube and for a selected number of the calculated random
distributions of defects.}
\label{GvsN}
\end{figure}

For each particular distribution of defects we can calculate the conductance as a function of energy
according to (\ref{G_temp_finita}). In Figure \ref{GvsE} we plot the conductance around a small energy region
around the Fermi level for the $(5,5)$ nanotube
and a mean distance between defects of \textsl{d} = 10.1 nm for different number of defects included in the nanotube. 
The nanotube conductance presents strong fluctuations,
with a maximum value close to the conductance quantum and
a minimum that becomes smaller for an increasing number of divacancies.
In fact, conductance never crosses the critical value of one conductance quantum once a certain number of
defects have been included. This result can be very clearly seen in Figure \ref{GvsN}, where
the conductance never reaches values higher than 1 G$_0$ after N = 5 \cite{Biel2005}. 
To clarify this issue, we have diagonalized the transmission matrix \cite{Jelinek2003}
to calculate the conductance for each one of the two conduction channels existing for energies
around the Fermi level. According to this, only one of the two conduction channels is contributing to the
total conductance once a number of 5-9 divacancies are presented in the nanotube. The origin of this
channel suppression remains, however, unclear.

The appropriately averaged resistance --according to the theory of disordered
one-dimensional systems \cite{Pendry}-- for a
statistically significant number of random configurations of defects
(up to 100 in the case of zero temperature, and up to 15 for the room temperature case,
where the dispersion with respect to the mean value is much smaller)
has been calculated for mean distances \textsl{d} between
10.1 nm and 75.5 nm. 

\begin{figure}[htbp]
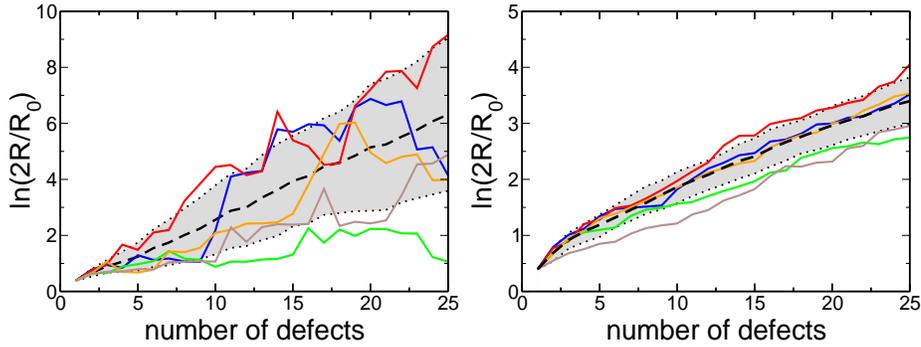

\vskip0.5cm
\centering
\includegraphics[width=6cm,height=4.5cm]{fig11_left.eps}
\includegraphics[width=6cm,height=4.5cm]{fig11_right.eps}
\caption{(color online) $(10,10)$ CNT: Colored lines:  $ln(2R/R_0)$ as a function of the number of defects,
N, for \textsl{d} = 75.5 nm and for a selected number of the calculated random
distributions of defects at T = 0 (left) and T = 300 K (right). Thick black lines: averaged resistance
for the total number of defect realizations considered. Grey shadowed region: mean quadratic deviation
with respect to the averaged resistance value. R$_0$ is the inverse of the conductance quantum G$_0$.}
\label{T0vsTamb}
\end{figure}

Figure \ref{T0vsTamb} shows the resistance, for the $(10,10)$ CNT, of a selected number of the calculated random
distributions of defects at both zero and room temperature. Curves with the same color in (left)
and (right) plots correspond to the same configuration of defects.
At zero temperature, the resistance for each defect realization strongly fluctuates
along the nanotube length, showing the characteristic behavior of disordered 1D system.
The average resistance, however, presents a clear exponential behavior as a function
of the length, which is a typical feature of strongly localized systems.
At room temperature, resistance fluctuations disappear for increasing values of T, leading
to very smooth curves when a critical temperature (scaling with ~1/d) is reached, but preserving
the exponential behavior. This fact confirms the presence of the Anderson localization regime
at room temperature for the defected nanotubes studied. This result has been confirmed by experimental
data \cite{Gomez2005,Flores2007}.
\begin{figure}[htbp]
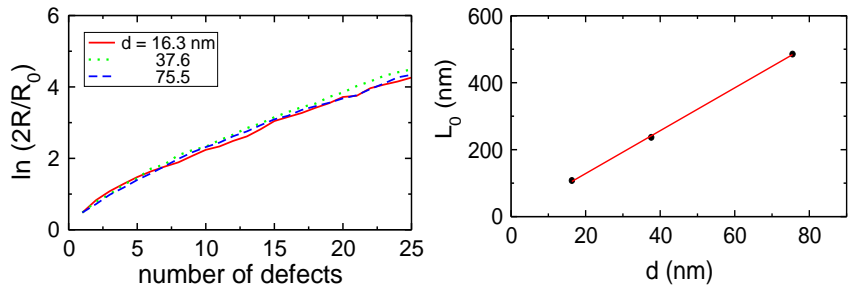

\vskip0.5cm
\centering
\includegraphics[width=5.5cm,height=3.7cm]{fig12_left.eps}
\includegraphics[width=5.5cm,height=3.7cm]{fig12_right.eps}
\caption{(color online) (left) Averaged $ln(2R/R_0)$ for different mean distances between defects \textsl{d} at
room temperature for the $(7,7)$ nanotube. (right)  Values for L$_0$, calculated by means of (\ref{ajuste_teoria}),
as a function of \textsl{d} for the same nanotube.}
\label{7_7Tamb}
\end{figure}

Figure \ref{7_7Tamb} shows the averaged resistance for different
mean distances at
room temperature for the $(7,7)$ nanotube. For all defect densities studied we have found
the strongly localized regime and the exponential increase of the resistance as a function of the
tube length. We can then fit
the average resistance for every \textsl{d} to the expression
\beq
R(L)\,=\,\frac{1}{2}\,R_0\,e^{L/L_0}\quad ,
\label{ajuste_teoria}
\eeq
where L is the length of the nanotube calculated as L = N$\cdot$ \textsl{d} and L$_0$ is
the localization length.
In this way we can calculate the localization length L$_0$ for every defect density.
These results for the $(7,7)$ nanotube are summarized in Figure \ref{7_7Tamb} (right).
A detailed analysis of the methodology and theoretical results obtained
at different temperatures, as well as the experimental data supporting these conclusions,
can be found in references \cite{Gomez2005,Biel2005} for
the  $(10,10)$ and in \cite{Flores2007} for the $(5,5)$.
\section{Conclusions}
In summary, we have analyzed, by means of a combination of
\textsl{ab initio} simulations and linear-scaling Green's
functions techniques, transport properties of armchair $(5,5)$, $(7,7)$ and $(10,10)$
carbon nanotubes with realistic disorder, namely mono- and di-vacancies.
In a first step, first-principles simulations for each defect have been performed
in order to determine the most stable reconstructions. The conductance profile
for every defect has been then analyzed. Our results show that reconstructed monovacancies present
an almost negligible drop of the nanotube conductance at the Fermi level compared to that of divacancies.
As the lateral orientation of the divacancy is found to be about
2 eV more stable than the vertical one
for all the nanotubes considered, only lateral divacancies have been
included in our analysis of long tubes with different defect densities.
Using an O(N) recursive procedure based on standard Green's functions techniques we have calculated
the localization length for several defect densities,
predicting the existence of the Anderson localization regime for defected carbon nanotubes at room temperature.
\ack
This work was partially supported by Spanish MCyT under contracts MAT2004-01271,
MAT2005-01298, MAT2002-01534 and NAN2004-09183-C10,
and by Comunidad de Madrid under contract S0505/MAT-0303 and the
European Community IST-2001-38052 and NMP4-CT-2004-500198 grants. A.R. acknowledges
support by the European CommunityNetwork of Excellence Nanoquanta
(NMP4-CT-2004-500198),SANES (NMP4-CT-2006-017310), DNA-NANODEVICES
(IST-2006-029192), NANO-ERA-Chemistry projects, Spanish MEC
(FIS2007-65702-C02-01), Basque Country University (SGIker ARINA)
and Basque Country Government (Grupos Consolidados 2007).
%

%%%%%%%%%%%%%%%%%%%%%%%%%%%%%%%%%%%%%%%%

%
\end{document}